\documentclass[journal]{IEEEtran}
\usepackage{generic}
\usepackage{cite}
\usepackage{amsmath,amssymb,amsfonts}
\usepackage{algorithmic}
\usepackage{graphicx}
\usepackage{textcomp}

\usepackage{hyperref}
\usepackage{subcaption}
\usepackage{array, booktabs}
\usepackage{balance}
\newcolumntype{P}[1]{>{\centering\arraybackslash}p{#1}}
\newcolumntype{M}[1]{>{\centering\arraybackslash}m{#1}}
\usepackage{adjustbox}
\usepackage{multirow}
\usepackage{caption}
\usepackage{float}
\usepackage{threeparttable}
\usepackage{xcolor}

\def\BibTeX{{\rm B\kern-.05em{\sc i\kern-.025em b}\kern-.08em
    T\kern-.1667em\lower.7ex\hbox{E}\kern-.125emX}}
\newcommand{\etal}{\emph{et~al.}}
\DeclareMathOperator{\bn}{bn}

\usepackage{tikz}
\usepackage{textcomp}
\usepackage{hyperref}

\newcommand\copyrighttext{%
  \footnotesize \textcopyright 2022 IEEE. This article has been accepted for publication in IEEE JOURNAL OF BIOMEDICAL AND HEALTH INFORMATICS. See \url{http://www.ieee.org/publications_standards/publications/rights/index.html} for copyright information.}
\newcommand\copyrightnotice{%
\begin{tikzpicture}[remember picture,overlay]
\node[anchor=south,yshift=10pt] at (current page.south) {\fbox{\parbox{\dimexpr\textwidth-\fboxsep-\fboxrule\relax}{\copyrighttext}}};
\end{tikzpicture}%
}

\begin{document}

\title{Anatomy-XNet: An Anatomy Aware Convolutional Neural Network for Thoracic Disease Classification in Chest X-rays}

\author{Uday Kamal, Mohammad Zunaed, Nusrat Binta Nizam,
        and Taufiq Hasan, \IEEEmembership{Senior Member, IEEE}
\thanks{Uday Kamal was with the mHealth Laboratory, Department of Biomedical Engineering, Bangladesh University of Engineering and Technology, Dhaka-1205, Bangladesh. He is now with the School of Electrical and Computer Engineering, Georgia Institute of Technology, Atlanta, GA 30332-0250 USA. This work was done while he was affiliated with mHealth Laboratory. (email: uday.kamal@gatech.edu).} 
\thanks{Mohammad Zunaed, Nusrat Binta Nizam, and Taufiq Hasan are with the mHealth Laboratory, Department of Biomedical Engineering, Bangladesh University of Engineering and Technology, Dhaka-1205, Bangladesh (e-mail: rafizunaed@gmail.com, nusratbintanizam@bme.buet.ac.bd, taufiq@bme.buet.ac.bd).}
}

\maketitle
\copyrightnotice

\begin{abstract}
Thoracic disease detection from chest radiographs using deep learning methods has been an active area of research in the last decade. Most previous methods attempt to focus on the diseased organs of the image by identifying spatial regions responsible for significant contributions to the model’s prediction. In contrast, expert radiologists first locate the prominent anatomical structures before determining if those regions are anomalous. Therefore, integrating anatomical knowledge within deep learning models could bring substantial improvement in automatic disease classification. Motivated by this, we propose Anatomy-XNet, an anatomy-aware attention-based thoracic disease classification network that prioritizes the spatial features guided by the pre-identified anatomy regions. We adopt a semi-supervised learning method by utilizing available small-scale organ-level annotations to locate the anatomy regions in large-scale datasets where the organ-level annotations are absent. The proposed Anatomy-XNet uses the pre-trained DenseNet-121 as the backbone network with two corresponding structured modules, the Anatomy Aware Attention (A\textsuperscript{3}) and Probabilistic Weighted Average Pooling (PWAP), in a cohesive framework for anatomical attention learning. We experimentally show that our proposed method sets a new state-of-the-art benchmark by achieving an AUC score of 85.78\%, 92.07\%, and, 84.04\% on three publicly available large-scale CXR datasets--NIH, Stanford CheXpert, and MIMIC-CXR, respectively. This not only proves the efficacy of utilizing the anatomy segmentation knowledge to improve the thoracic disease classification but also demonstrates the generalizability of the proposed framework.
\end{abstract}


\section{Introduction}
\IEEEPARstart{C}{hest} radiography (CXR) is the most commonly used primary screening tool for assessing thoracic diseases \cite{raoof2012interpretation}. Each year a massive number of CXRs are produced, and the diagnosis is performed mainly by radiologists. With the severe shortage of expert radiologists, especially in developing countries, computer-aided disease detection from chest radiographs is considered the future of medical diagnosis \cite{yu2011automatic, jaeger2013automatic}. Advancement in deep learning and artificial intelligence offers several ways of rapid, accurate, and reliable screening techniques \cite{litjens2017survey}. These techniques can significantly impact the health systems in the resource-constrained regions of the world where there is a high prevalence of thoracic diseases and a shortage of expert radiologists. \par

\begin{figure}[t]
    \centering
	\includegraphics[width=0.9\linewidth]{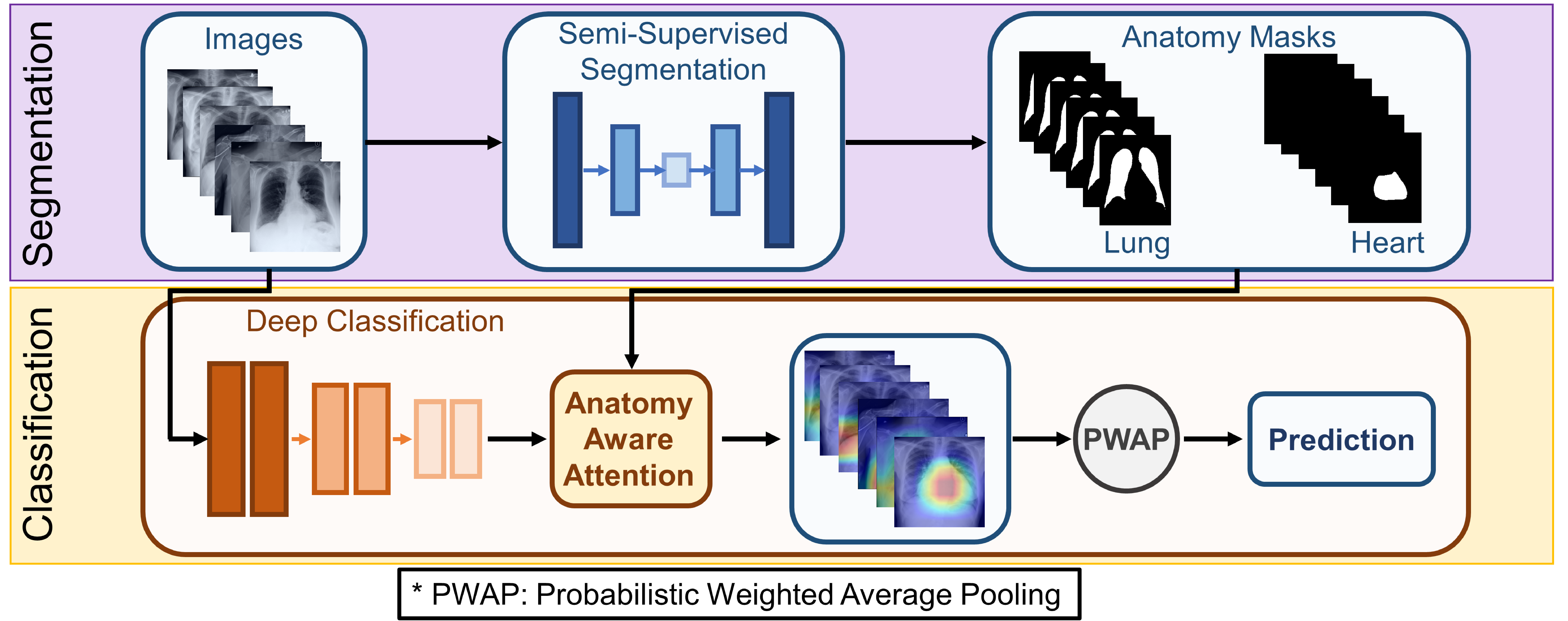}
    \caption{Overview of the proposed semi-supervised anatomy-aware attention-based thoracic disease classification framework. A semi-supervised technique is utilized to generate anatomy masks for unannotated CXR images. Then, with the help of our proposed novel anatomy-aware attention module, anatomical information is integrated into the classification network for pathology detection.}
    \label{overall_framework}
\end{figure}
Driven by many publicly accessible large-scale CXR datasets, a significant amount of research efforts have been carried out for the automatic diagnosis of thoracic diseases. Wang \etal \cite{8099852} first announced the ChestX-ray14 dataset and proposed a unified weakly-supervised classification network by introducing various multi-label DCNN losses based on ImageNet pre-trained deep CNN models. LLAGnet \cite{Chen2020LesionLA} is a novel lesion location attention guided network containing two corresponding attention modules which focus on the discriminative features from lesion areas for multi-label thoracic disease classification in CXRs. Wang \etal \cite{wang2020triple} proposed a DenseNet-121 based triple learning approach that integrates three attention modules which are unified for channel-wise, element-wise, and scale-wise attention learning. \par
In medical practice, interpretation of chest X-rays, or any other medical imaging modalities for that matter, requires an understanding of the relevant human anatomy that is being imaged. For example, fundamental analysis of chest X-rays involves the radiologist determining if the trachea is central, the lungs are uniformly expanded, the lung fields are clear, and the heart size is normal \cite{clarke2017chest}. These and other similar observations form the basis of CXR interpretation by human vision, where it is clear that knowledge of anatomical structures is vital. Real-world radiologists tend to locate the vital anatomy regions first and then determine if those regions have abnormalities. Similarly, successful implementation of deep learning-based thoracic disease classification approaches requires not only higher accuracy but also interpretability. However, most previous research works in automated analysis of CXRs do not consider this aspect and address the problem as any other computer vision problem. Most previous methods employed a global learning strategy \cite{8099852, Rajpurkar2017CheXNetRP}, or relied on attention mechanisms \cite{Tang2018AttentionGuidedCL, Chen2020LesionLA, Yan2018WeaklySD}, that try to determine the spatial regions that are more responsible for model prediction. In \cite{XU202196, Keidar2021, liu2019sdfn, ARIASGARZON2021100138} methods have been proposed to integrate segmentation masks into the backbone framework. However, proper contour-level annotations for large-scale datasets \cite{8099852, irvin2019chexpert, johnson2019mimiccxrjpg} are unavailable. Generating segmentation masks from a minimal amount of annotated datasets (e.g., Japanese Society of Radiological Technology (JSRT) \cite{Shiraishi2000DevelopmentOA}) for these large-scale datasets lead to imperfect segmentation masks. However, the approaches in \cite{XU202196, Keidar2021, liu2019sdfn, ARIASGARZON2021100138} did not consider the effect of the imperfect segmentation masks in their proposed frameworks. These imperfections of the segmentation masks lead to difficulty for the backbone model to properly identify the anatomy regions. \par
In this work, we propose an anatomy-aware attention-based architecture named Anatomy-XNet that utilizes the anatomy segmentation information along with CXRs frames to classify thoracic diseases. A significant challenge to integrate anatomy information into the framework is the lack of proper contour-level anatomy region annotations for large-scale datasets such as NIH \cite{8099852}, CheXpert \cite{irvin2019chexpert}, and MIMIC-CXR \cite{johnson2019mimiccxrjpg}. To solve this problem, we leverage a semi-supervised learning technique \cite{Mondal2019RevisitingCF}, requiring only a handful of annotated instances that enables us to utilize small scale dataset like JSRT \cite{Shiraishi2000DevelopmentOA} to train the segmentation network and generate the anatomy segmentation masks for the NIH, CheXpert, and MIMIC-CXR datasets. However, one downside of this method is that it doesn't guarantee similar performance compared to any supervised learning method \cite{Mondal2019RevisitingCF}. In order to mitigate this problem, we incorporate a novel structured module called Anatomy Aware Attention (A\textsuperscript{3}) on top of the backbone feature extractor, Densenet-121, in a united framework. The A\textsuperscript{3} module not only reinforces the sensitivity of the different stages of the model to prioritize the anatomical location responsible for a thoracic disease, but also retains information outside the masks through the residual attention vector and thus is less affected by the imperfect anatomy masks. In addition, we propose a novel pooling operation layer, named Probabilistic Weighted Average Pooling (PWAP), that explicitly leverages the probability attention map derived from the feature activation map to enhance the salient regions of the feature space. An overview of our proposed framework is presented in Fig.~\ref{overall_framework}. The contributions of this paper are summarized as follows:
\begin{itemize}
  \item We propose novel hierarchical feature-fusion-based A\textsuperscript{3} modules that learn to re-calibrate the feature maps in different stages of the model based on anatomical knowledge to improve the classification performance and the model's robustness to imperfection in anatomy masks.
  \item We incorporate novel PWAP modules that utilizes a learnable re-weighting mechanism based on spatial feature importance before performing spatial feature aggregation.
  \item Our proposed Anatomy-XNet achieves new state-of-the-art performances with AUC scores of 85.78\%, 92.07\%, and, 84.04\% on three publicly available large-scale CXR datasets, NIH, Stanford CheXpert, and, MIMIC-CXR, respectively. These extensive experiments demonstrate the effectiveness of utilizing prior anatomy knowledge and prove the generalizability of the proposed framework.
\end{itemize}

\section{Related work}
\subsection{Organ Segmentation from Chest Radiographs}
There are several methods for organ segmentation from a CXR image. Among the classical signal processing based methods, a hybrid approach by Shao \etal \cite{6737258} combining active shape and appearance models, a combined approach of landmark-based segmentation and a random forest classifier by Ibragimov \etal \cite{7493451}, an active shape framework addressing the initialization dependency of these active shape models proposed by Xu \etal \cite{XU2012452} are noteworthy. In the advent of deep learning, Convolutional neural network (CNN) based segmentation of medical images has attracted wider attention of researchers. An end-to-end contour-aware CNN-based segmentation method is shown to provide organ contour information and improve the segmentation accuracy \cite{kholiavchenko2020contour}. In \cite{munawar2020segmentation}, lung segmentation is performed from CXRs using Generative adversarial networks. However, this model is not generalizable to new datasets. Two-stage deep learning techniques such as patch classification and reconstruction of lung fields can be used for lung segmentation from CXR images \cite{souza2019automatic}.
    
\copyrightnotice
\subsection{Disease Classification from Chest Radiographs}
\subsubsection{Methods without Utilizing Segmentation Masks}
\label{classification_literature}
Many signal processing and deep learning approaches have been proposed to classify thoracic diseases in recent years. Tang \etal \cite{Tang2018AttentionGuidedCL} identified the disease category and localized the lesion areas through an attention-guided curriculum learning method. In \cite{ho2019multiple}, multiple feature integration is presented using shallow handcrafted techniques and a pre-trained deep CNN model. DualCheXNet \cite{chen2019dualchexnet} is an approach that enables two different feature fusion operations, such as feature-level fusion and decision level fusion, which form the complementary feature learning embedded in the network. LLAGnet \cite{Chen2020LesionLA} is a novel lesion location attention guided network containing two corresponding attention modules which focus on the discriminative features from lesion locations for multi-label thoracic disease classification in CXRs. 
\begin{figure*}[!t]
	\centering
	\includegraphics[width=0.9\linewidth]{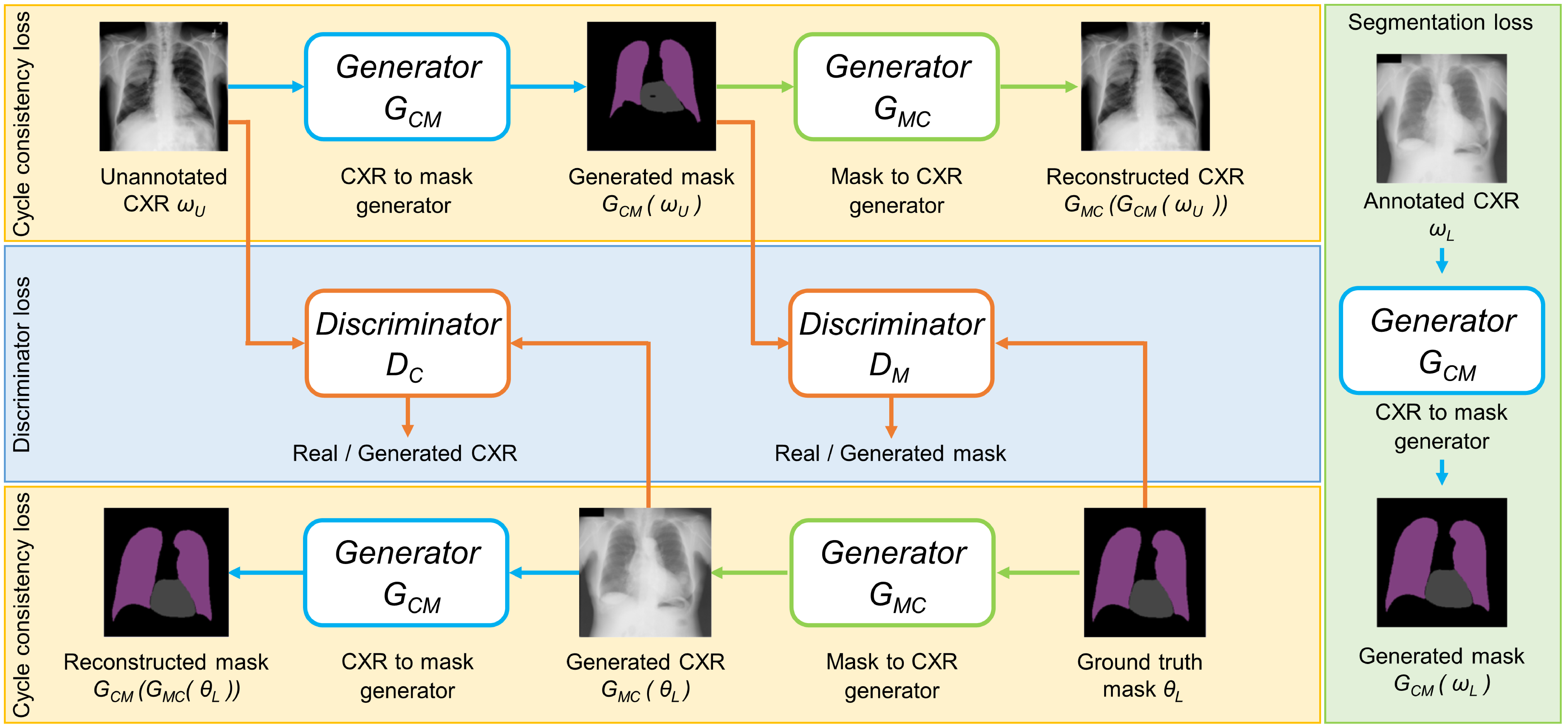}
	\caption{Overview of the semi-supervised anatomy segmentation architecture.}
	\label{cyclegan_archi}
\end{figure*}
Guan \etal \cite{GUAN2020259} proposed a category-wise residual attention learning framework for multi-label thoracic disease classification. Rajpurkar \etal \cite{Rajpurkar2017CheXNetRP} exploited a modified 121-layer DenseNet named CheXNet, for diagnosis of all 14 pathologies in the ChestXray14 dataset, especially for pneumonia. In \cite{wang2020triple}, a triple learning approach integrating a unified channel-wise, element-wise, and scale-wise attention modules are used. They can simultaneously learn disease-discriminative channels, locations, and scales for effective diagnosis. Hou \etal \cite{Hou2021MultiLabelLW} fused semantic features from radiology reports along with encoded X-ray features to feed into transformer encoder to utilize both CXR images and metadata related to them. Zhang \etal \cite{Zhang_Wang_Xu_Yu_Yuille_Xu_2020} proposed a medical concept graph, based on prior knowledge, to diagnose CXR images. Seyyed-Kalantar \etal \cite{SeyyedKalantari2021CheXclusionFG} examined the extent to which state-of-the-art deep learning classifiers show true positive rate disparity among different protected attributes. Allaouzi \etal \cite{8719904} explored binary relevance (BR), label powerset, and classifier chain in terms of label dependencies. Yan \etal \cite{Yan2018WeaklySD} proposed a weakly supervised deep learning framework equipped with squeeze and excitation blocks, multi-map transfer, and max-min pooling for classifying and localizing suspicious lesion regions. Luo \etal \cite{Luo2020DeepME} adopted task-specific adversarial training and an uncertainty-aware temporal ensemble of model predictions to address the domain and label discrepancies across different datasets. To handle label uncertainty on the CheXpert dataset, Irvin \etal \cite{irvin2019chexpert} trained a DenseNet-121 on CheXpert with various labeling policies such as U-Ignore, U-Ones, and U-Zeros policies. Pham \etal \cite{PHAM2021186} exploited dependencies among abnormality labels and utilized label smoothing technique for better handling of uncertain samples in the CheXpert dataset. However, a systematic exploration of the potential of integrating anatomical prior to improve the classification performance was absent in all the above mentioned methods. 

\subsubsection{Methods Utilizing Segmentation Masks}
\label{classification_literature_with_segmentation_masks}
Xu \etal \cite{XU202196} proposed a dual-stage approach (segmentation and classification) to utilize mask-attention-mechanism as spatial attention to adjust salient features of the CNN. Their attention mask suppresses the receptive field of the CNN based on their overlapping rates with the segmentation masks. Keidar \etal \cite{Keidar2021} proposed a deep learning-based model, along with the segmentation masks as additional input, for the detection of COVID-19 from CXRs. Segmentation-based Deep Fusion Network (SDFN) \cite{liu2019sdfn} is a method that leverages the domain knowledge and the higher-resolution information of local lung regions. The local lung regions are identified using Lung Region Generator, and discriminative features are extracted using two CNN models. Then these features are fused by the feature fusion module for the disease classification process. Arias-Garzón \etal \cite{ARIASGARZON2021100138} proposed a two-stage method where the surrounding area around anatomy regions are removed from the CXR image based on the segmentation masks to remove any classification bias towards the extraneous (i.e., non-anatomy) regions of the image. Afterward, they fed the CXR image constrained by the segmentation mask to a CNN model. Overall, the methods described in \cite{XU202196, Keidar2021, liu2019sdfn, ARIASGARZON2021100138} utilized small-scale annotated datasets in a supervised training setting to generate segmentation masks for large-scale datasets used in their approaches. However, in these methods, the effect of imperfect segmentation masks was not considered, which naturally arises from supervised training of the segmentation network using out-of-distribution data resources.
\copyrightnotice

\section{Methodology}
\subsection{Semi-supervised Anatomy Segmentation Network}
For semi-supervised segmentation of anatomy regions, we adopted the method from \cite{Mondal2019RevisitingCF} which is based on the popular CycleGAN architecture \cite{cyclegan}. The CycleGAN architecture comprises of four interconnected blocks, two conditional generators, and two discriminators as illustrated in Fig.~\ref{cyclegan_archi}. The first generator ($G_{CM}$), corresponding to the segmentation network that we want to obtain, learns a mapping from a CXR image to its anatomy segmentation mask. The first discriminator ($D_M$) takes either the generated mask from $G_{CM}$ or the real segmentation mask as input, and learns to differentiate one from another. Conversely, the second generator ($G_{MC}$) learns to map a segmentation mask back to its CXR image. The second discriminator ($D_C$) receives a CXR image as input (either a real CXR image or a generated CXR from $G_{MC}$) and predicts whether this image is real or generated. To enforce cycle consistency criterion, the segmentation network is trained in a way so that feeding the segmentation mask generated by $G_{CM}$ for a CXR image into $G_{MC}$ returns the same CXR image. Similarly, passing back the CXR image generated by $G_{MC}$ to $G_{CM}$ for a segmentation mask returns the same mask.

\subsubsection{\textbf{Loss functions}}
The segmentation setting contains two distinct subsets: subset $L$, containing annotated CXR images $\boldsymbol{\omega}_L$ and their corresponding ground-truth masks $\boldsymbol{\theta}_L$, and subset $U$, which contains unannotated CXR images $\boldsymbol{\omega}_U$. We train the generator module $G_{CM}$ to generate segmentation mask by imposing the following loss function,
\begin{align}
\mathcal{L}_{gen}^{M}(G_{CM})&= \mathbb{E}_{\boldsymbol{\omega},\boldsymbol{\theta} \in \boldsymbol{\omega}_{L},\boldsymbol{\theta}_{L}}\big[\mathcal{H}(\boldsymbol{\theta},G_{CM}(\boldsymbol{\omega}))\big] \\
\mathcal{H}(\boldsymbol{\theta},\Tilde{\boldsymbol{\theta}}) &= -\sum_{j=1}^{N}\sum_{k=1}^{K} \boldsymbol{\theta}_{j,k}\log \Tilde{\boldsymbol{\theta}}_{j,k}
\end{align}
Here, $\mathcal{H}$ is the pixel-wise cross-entropy, $\boldsymbol{\theta}_{j,k}$ and $\Tilde{\boldsymbol{\theta}}_{j,k}$ are the annotated segmentation mask and predicted probabilities that pixel j $\in \{1,...,N\} $ has label k $\in \{1,...,K\} $. We employ a pixel-wise L2 norm between an annotated CXR and the CXR generated from its corresponding segmentation mask as a supervised loss to train the CXR generator $G_{MC}$:
\begin{equation}
\mathcal{L}_{gen}^{C}(G_{MC})= \mathbb{E}_{\boldsymbol{\omega},\boldsymbol{\theta} \in \boldsymbol{\omega}_{L},\boldsymbol{\theta}_{L}}\big[{\Vert G_{MC}(\boldsymbol{\theta})-\boldsymbol{\omega} \Vert}_{2}^{2}\big]
\end{equation}
Two additional losses, adversarial and cycle consistency losses, are incorporated to exploit unannotated CXR images. We use the adversarial losses to train the generators and discriminators in a competing fashion and help the generators produce realistic CXR image and anatomy segmentation mask. Suppose that $D_M(\boldsymbol{\theta})$ is the predicted probability that segmentation mask $\boldsymbol{\theta}$ correspond to an annotated CXR's segmentation mask. We define the adversarial loss for $D_M$ as,
\begin{align}
\mathcal{L}_{disc}^{M}(G_{CM},D_M) =    \ &\mathbb{E}_{\boldsymbol{\theta} \in \boldsymbol{\theta}_{L}}\big[(D_{M}(\boldsymbol{\theta})-1)^{2}\big]+\nonumber\\
& \mathbb{E}_{\boldsymbol{\omega}^{'}\in \boldsymbol{\omega}_{U}}\big[(D_{M}(G_{CM}(\boldsymbol{\omega}^{'})))^2\big]
\end{align}
Let $D_{C}(\boldsymbol{\omega})$ be the predicted probability that a CXR $\boldsymbol{\omega}$ is real. We get the adversarial loss for the CXR discriminator by,
\begin{align}
\mathcal{L}_{disc}^{C}(G_{MC},D_C)= \
&\mathbb{E}_{\boldsymbol{\omega}^{'} \in \boldsymbol{\omega}_{U}}\big[ (D_{C}(\boldsymbol{\omega}^{'})-1)^{2}\big]+\nonumber\\
&\mathbb{E}_{\boldsymbol{\theta} \in \boldsymbol{\theta}_{L}}\big[(D_{C}(G_{MC}(\boldsymbol{\theta})))^2 \big]
\end{align}
The first cycle consistency loss measures the difference between an unannotated CXR and the regenerated CXR after passing through generators $G_{CM}$ and $G_{MC}$ sequentially.
\begin{equation}
\mathcal{L}_{cycle}^{C}(G_{CM},G_{MC})= 
\mathbb{E}_{\boldsymbol{\omega}^{'}\in \boldsymbol{\omega}_{U}}\big[\Vert G_{MC}(G_{CM}(\boldsymbol{\omega}^{'}))-\boldsymbol{\omega}^{'} \Vert_{1}\big]
\end{equation}
We use cross-entropy to evaluate the difference between an annotated and regenerated segmentation mask after passing through generators $G_{MC}$ and $G_{CM}$ in sequence:
\begin{equation}
\mathcal{L}_{cycle}^{M}(G_{CM},G_{MC})= 
\mathbb{E}_{\boldsymbol{\theta} \in \boldsymbol{\theta}_{L}}\big[ \mathcal{H}(\boldsymbol{\theta},G_{CM}(G_{MC}(\boldsymbol{\theta})))\big]
\end{equation}
Finally, the total loss is obtained by combining all loss terms:
\begin{equation}
\mathcal{L}_{total}=\mathcal{L}_{gen}^{M}+\mathcal{L}_{gen}^{C}+\mathcal{L}_{cycle}^{M}+\mathcal{L}_{cycle}^{C}
-\mathcal{L}_{disc}^{M}-\mathcal{L}_{disc}^{C}
\end{equation}
We perform the learning in an alternating fashion. The parameters of the generators are optimized while considering those of the discriminators as fixed and vice versa.
\begin{figure*}[!t]
	\centering
	\includegraphics[width=0.9\linewidth]{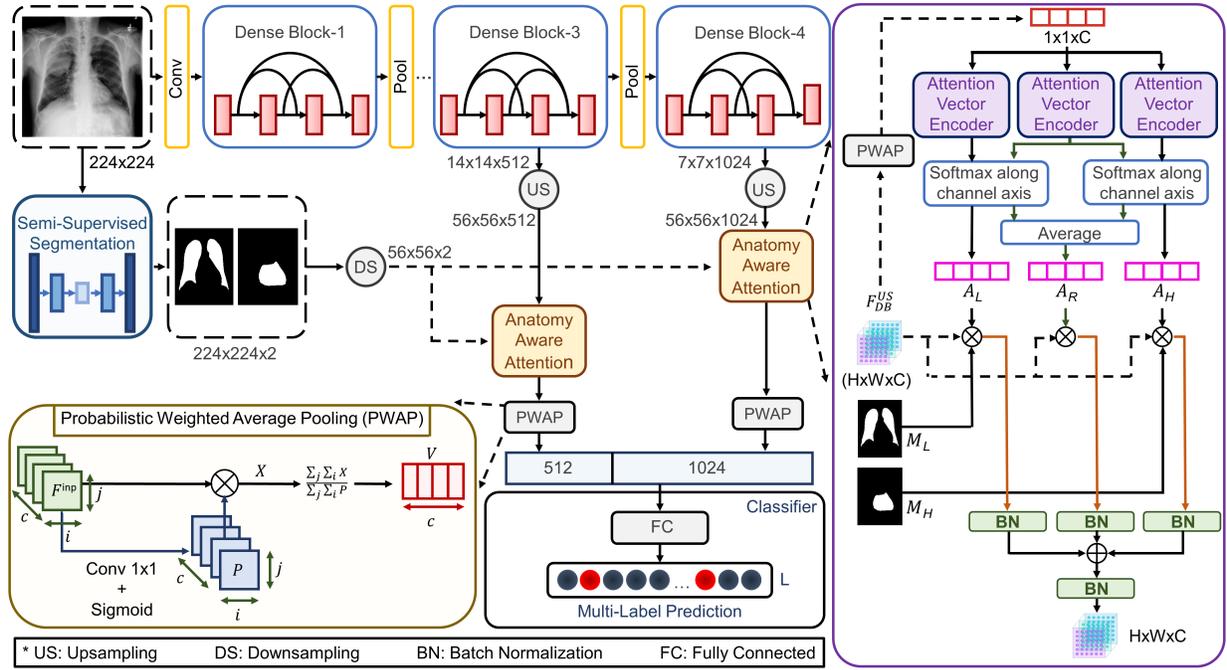}
 	\caption{The architecture of the proposed Anatomy-XNet. The anatomy-aware attention (A\textsuperscript{3}) modules operate on the upsampled feature spaces from dense block-3 and dense block-4 with the help of downsampled anatomy masks. These anatomy masks are derived from the segmentation network in a semi-supervised manner. The feature spaces calibrated with the supervision of anatomy knowledge from each of the A\textsuperscript{3} modules are pooled by the PWAP layers and concatenated. The classifier module outputs pathology class scores by taking these concatenated pooled features as input.}
 	\label{anatomyxnet_archi}
\end{figure*}
\subsection{Anatomy-XNet}
The proposed Anatomy-XNet architecture is illustrated in Fig.~\ref{anatomyxnet_archi}. We utilize transfer learning on DenseNet-121 \cite{8099726} architecture pre-trained on the ImageNet and use it as our backbone model. The A\textsuperscript{3} modules operate on the high-level feature space encoded by the dense blocks (DB) to enforce attention supervision guided by the anatomy masks. We perform downsampling and upsampling on the anatomy masks and feature space to an intermediate shape before passing them to an A\textsuperscript{3} module. The different components of the proposed Anatomy-XNet are described in the following sub-sections.
\subsubsection{\textbf{Probabilistic Weighted Average Pooling (PWAP) Module}} \label{sec_PWAP}
The traditional global average pooling or max-pooling layer provides the same weight to all spatial regions of the input. However, in many cases, the object of interest may reside in a salient region that is more important than others. Usually, thoracic diseases are often characterized by an anatomy region and lesion areas that constitute much smaller portions than the entire image. Thus, to further enhance the attention mechanism, we use a PWAP module in conjunction with the A\textsuperscript{3} block and a PWAP module within the A\textsuperscript{3} block. This module explicitly leverages the probability attention map, derived from the input feature activation space, to enhance the most discriminative regions of the feature map before applying the pooling operation. In this module, we learn the weight of each spatial position to guide Anatomy-XNet towards lesion localization during training through a 1\texttimes1 convolutional filter. This 1\texttimes1 filter has been chosen as we aim to learn the weight at a single spatial position; surrounding information is unwanted. First, we get the probability map $\mathbf{P} \in \mathbb{R}^{H \times W \times 1}$ from a input feature map $\mathbf{F}^{inp} \in \mathbb{R}^{H \times W \times C}$ by,
\begin{equation}
\mathbf{P}_{i,j,1} = \mathcal{S}\left(\sum_{c=1}^{C}\mathbf{K}_{1,1,c}*\mathbf{F}^{inp}_{i,j,c}\right)
\end{equation}
Here, $i \in \{1,...,H\}, \ j \in \{1....,W\}$, $\mathcal{S}(\cdot)$ denotes the sigmoid function, and $\mathbf{K}$ is the learnable convolutional filter. Afterward, we elementwise multiply the probability map $\mathbf{P}$ with the input feature map $\mathbf{F}^{inp}$ to obtain the weighted feature space $\mathbf{X}$. Then we normalize the feature space $\mathbf{X}$ and finally, obtain the pooled feature vector $\mathbf{V}$ of size 1$\times$1$\times$C by,
\begin{align}
 \mathbf{X}_{i,j,c} &  = \mathbf{F}^{inp}_{i,j,c} \odot \mathbf{P}_{i,j,c} \\
 \mathbf{V}_{1,1,c} &  = \frac{\sum_{i=1}^{H}\sum_{j=1}^{W}\mathbf{X}_{i,j,c}}{\sum_{i=1}^{H}\sum_{j=1}^{W}\mathbf{P}_{i,j,c}}
\end{align}
\copyrightnotice

\subsubsection{\textbf{Anatomy Aware Attention (\texorpdfstring{A\textsuperscript{3}}{A3}) Module}}
The Anatomy-XNet consists of two A\textsuperscript{3} modules. The first A\textsuperscript{3} module is connected to the third dense block (DB-3), and the other A\textsuperscript{3} module works with the fourth dense block (DB-4). Each one of them takes the upsampled high-level feature map $\mathbf{F}_{DB}^{US} \in \mathbb{R}^{H \times W \times C}$ generated by their corresponding DB block, and the downsampled anatomy segmentation masks $\mathbf{M} \in \mathbb{R}^{H \times W \times 2}$ as inputs. Using a PWAP module, the feature map is pooled to a feature vector $\mathbf{V} \in \mathbb{R}^{1 \times 1 \times C } $. This feature vector $\mathbf{V}$ is then passed through three different Attention Vector Encoder (AVE) modules to get the three feature vectors $\mathbf{A}_{1}$, $\mathbf{A}_{2},\ \mathbf{A}_{3} \in \mathbb{R}^{1 \times 1 \times C}$. The detailed architecture of an AVE module is described in Table~\ref{attn_vct_encoder_structure_table}. The architecture is designed to introduce the bottleneck mechanism in the AVE module, which is inspired by the Squeeze-and-Excitation block \cite{8578843}. To introduce bottleneck, feature vector $\mathbf{V}$ is first squeezed into dimension $1 \times 1 \times (C/r)$ and later excited back to $1 \times 1 \times C$. The value of C is 512 and 1024 for A\textsuperscript{3} modules connected to DB-3, and DB-4, respectively. In both cases, r is 0.5. \par
\copyrightnotice
We aim to give relevant importance to lung and heart anatomy compared to background regions. One straightforward way is to apply softmax operation across all the attention vectors to get that relevancy scores. But one drawback of this approach is that it will make the attention scores for lung and heart mask attention vectors dependent on each other. However, pathologies related to the heart are independent of whether the CXR contains lung pathologies or not. This motivates us to design the softmax operations across the attention vectors in such a way that the lung attention vector and heart attention vector are independent of each other, but the residual attention vector is jointly dependent on both of them. First, we apply softmax function between $\mathbf{A}_{1}$ and $\mathbf{A}_{2}$:
\begin{table}[b]
    \centering
    \caption{\textsc{Attention Vector Encoder Structure.}} 
        \begin{tabular}{ccc} 
        \toprule
        \bf Layer (Type) & \bf Input Shape & \bf Output Shape \\ 
        \hline
        \hline
        FC-1 (Fully Connected) & (C) & (C/r) \\
        ReLU-1 (ReLU) & (C/r) & (C/r) \\
        BN-1 (Batch Normalization) & (C/r) & (C/r)  \\
        FC-2 (Fully Connected) & (C/r) & (C) \\
        ReLU-2 (ReLU) & (C) & (C)  \\
        BN-2 (Batch Normalization) & (C) & (C)  \\
        \bottomrule
        \end{tabular}
    \label{attn_vct_encoder_structure_table}
        \begin{tabular}{cc}
            * C: Channel Dimension & r: Reduction ratio
        \end{tabular}
\end{table}
\begin{equation}
\sigma(\mathbf{A}_k)_i=\frac{\exp{(\mathbf{A}_k)_i}}{\sum_{j=1}^{2}{\exp{(\mathbf{A}_j)_i}}}\ ,  i \in \{1,...,C\}, k \in \{1,2\}
\end{equation}
Here, $\sigma(\cdot)$ represents softmax operation, $j$ and $k$ represent feature vector indices, and $i$ represents the $i^{th}$ channel value of a feature vector. Thus, we obtain two attention vectors where each feature value across the channel dimension depends on each other. We name these two attention vectors as the lung and lung-complementary attention vectors denoted by $\mathbf{A}_{L}$ and $\overline{\mathbf{A}_{L}}$, respectively. These quantities are related by,
\begin{equation}
\bigl(\mathbf{A}_{L}\bigr)_i+\bigl(\,\overline{\mathbf{A}_{L}}\,\bigr)_i=1,\ i \in \{1,...,C\} \label{eq_attention_1} 
\end{equation}
Similarly, we apply softmax on $\mathbf{A}_{2}$ and $\mathbf{A}_{3}$ by,
\begin{equation}
\sigma(\mathbf{A}_k)_i=\frac{\exp{(\mathbf{A}_k)_i}}{\sum_{j=2}^{3}{\exp{(\mathbf{A}_j)_i}}}\ ,  i \in \{1,...,C\}, k \in \{2,3\}
\end{equation}
Here, $j$ and $k$ represent feature vector indices, and $i$ represents the $i^{th}$ channel value of a feature vector. Similarly, we obtain two attention vectors where each feature value across the channel dimension depends on each other. We name these two attention vectors as the heart-complementary and heart attention vectors denoted by $\overline{\mathbf{A}_{H}}$ and $\mathbf{A}_{H}$, respectively. These quantities are related by,
\begin{equation}
\bigl(\mathbf{A}_{H}\bigr)_i+\bigl(\,\overline{\mathbf{A}_{H}}\,\bigr)_i=1,\ i \in \{1,...,C\} \label{eq_attention_2}
\end{equation}
The lung-complementary regions and heart-complementary regions have considerable overlap between them. For this reason, proper weighting between the lung-complementary and heart-complementary attention vectors is needed. Finally, we get the residual attention vector by,
\begin{equation}
\mathbf{A}_{R}=\alpha \overline{\mathbf{A}_{L}}+\beta \overline{\mathbf{A}_{H}}
\label{eq_attention_3}
\end{equation}
Here, the value of hyperparameters are: $\alpha$=0.5 and $\beta$=0.5, which are inferred from a grid search with a cross-validation. Next, we downsample and broadcast the lung and heart masks to the dimension of $H\times W \times C$ by repeating them in the channel ($C$) axis. We denote the downsampled and broadcasted lung and heart masks, respectively, as $\mathbf{M}_{L} \in \mathbb{R}^{H \times W \times C}$ and $\mathbf{M}_{H} \in \mathbb{R}^{H \times W \times C}$. Afterward, we element-wise multiply the attention vectors, $\mathbf{A}_{L}$ with $\mathbf{F}_{DB}^{US}$ and $\mathbf{M}_{L}$, $\mathbf{A}_{H}$ with $\mathbf{F}_{DB}^{US}$ and $\mathbf{M}_{H}$, and $\mathbf{A}_{R}$ with $\mathbf{F}_{DB}^{US}$ to get three feature spaces, $\mathbf{R}_{L}$, $\mathbf{R}_{H}$, and $\mathbf{R}_{R}$, respectively.
\begin{align}
 \mathbf{R}_{L} &= \mathbf{A}_{L} \odot \mathbf{M}_{L} \odot \mathbf{F}_{DB}^{US} \label{eqn_RL} \\
 \mathbf{R}_{H} &= \mathbf{A}_{H} \odot \mathbf{M}_{H} \odot \mathbf{F}_{DB}^{US} \\
 \mathbf{R}_{R} &= \mathbf{A}_{R} \odot \mathbf{F}_{DB}^{US} 
\end{align}
where $\odot$ represents the element-wise multiplication operation. Thus, we obtain two anatomy attentive feature space $\mathbf{R}_{L},\ \mathbf{R}_{H} \in \mathbb{R}^{H \times W \times C}$, and the residual attentive feature space, $\mathbf{R}_{R} \in \mathbb{R}^{H \times W \times C}$. For faster convergence and removal of any internal covariate shift among $\mathbf{R}_{L},\ \mathbf{R}_{H}$, and $\mathbf{R}_{R}$, batch normalization operation is applied individually. Next, we sum all of the three feature spaces and apply batch normalization to obtain the final feature space $\mathbf{R} \in \mathbb{R}^{H \times W \times C}$ by,
\begin{equation}
 \mathbf{R}=\bn\bigg(\bn(\mathbf{R}_{L})+\bn(\mathbf{R}_{H})+\bn(\mathbf{R}_{R})\bigg)
\end{equation}
Here, $\text{bn}(\cdot)$ denotes the batch normalization operation. Since $\mathbf{A}_{L}$ is multiplied by $\mathbf{M}_{L}$ and $\mathbf{F}_{DB}^{US}$, $\mathbf{A}_{L}$ provides attention to the spatial regions responsible for respiratory diseases. To verify this, let us define the loss function score $\mathcal{L}$ and take the gradient of $\mathcal{L}$ with respect to the lung attention vector $\mathbf{A}_{L}$.
\begin{align}
\frac{\partial \mathcal{L}}{\partial (A_{L})_{i,j}^k} 
 &=  \sum_{i}\sum_{j}\frac{\partial \mathcal{L}}{\partial (R_{L})_{i,j}^k} \cdot \frac{\partial (R_{L})_{i,j}^k}{\partial (A_{L})_{i,j}^k}
\end{align}
where, $A_{L} \in \mathbb{R}^{1 \times 1 \times C},\ R_{L} \in \mathbb{R}^{H \times W \times C},$ and $i \in \{1,2,...H\}, j \in \{1,2,...W\}, k \in \{1,2,...C\}$. From equation \eqref{eqn_RL} we get that, $\mathbf{R}_{L}=\mathbf{A}_{L} \odot \mathbf{M}_{L} \odot \mathbf{F}_{DB}^{US}$, where $M_{L} \in \mathbb{R}^{H \times W \times C}$. Hence,
\begin{align}
\frac{\partial \mathcal{L}}{\partial (A_{L})_{i,j}^k} 
&=  \sum_{i}\sum_{j}\frac{\partial \mathcal{L}}{\partial (R_{L})_{i,j}^k} \! \cdot \! \frac{\partial \left((A_{L})_{i,j}^k \! \cdot \! (M_{L})_{i,j}^k \! \cdot \! (F_{DB}^{US})_{i,j}^k\right)}{\partial (A_{L})_{i,j}^k} \nonumber \\
&=  \sum_{i}\sum_{j}(M_{L})_{i,j}^k \! \cdot \! \frac{\partial \left((A_{L})_{i,j}^k \! \cdot \! (F_{DB}^{US})_{i,j}^k\right)}{\partial (A_{L})_{i,j}^k} \! \cdot \! \frac{\partial \mathcal{L}}{\partial (R_{L})_{i,j}^k}
\end{align}
The value of $(M_{L})_{i,j}^k$ is 1 at any spatial position if it is the lung region, otherwise is 0. As a result, the gradient for the lung attention vector ($\mathbf{A}_{L}$) is weighted according to the lung-mask region. Similarly, the gradient for the heart attention vector ($\mathbf{A}_{H}$) is weighted according to the heart-mask region, making $\mathbf{A}_{H}$ to provide attention to the heart-related (cardiac) diseases. \par 
The residual feature space ($\mathbf{R}_{R}$) contains feature activation values responsible for the whole input feature map that includes predicted anatomy mask regions, as well as areas other than predicted anatomy mask regions. The areas other than the predicted anatomy mask regions include any left-out anatomy regions from the predicted segmentation masks. The residual attentive feature space ($\mathbf{R}_{R}$) is responsible for attention to these regions, enabling the Anatomy-XNet a relaxed view constraint on the imperfect segmentation masks and thus making it less affected by the imperfect anatomy masks.
\copyrightnotice
\subsubsection{\textbf{Classifier}} \label{sec_xnet_classifier}
Let $V_{1} \in \mathbb{R}^{1 \times 1 \times C_{1}}$, $V_{2} \in \mathbb{R}^{1 \times 1 \times C_{2}}$ be the pooled feature vectors from the PWAP modules connected to the A\textsuperscript{3} modules that work on the DB-3 and DB-4, respectively. Here, $C_{1}=512$ and $C_{2}=1024$. We concatenate $V_{1}$, $V_{2}$ together and pass them through a fully connected (FC) layer. The output $f_{i}^{I}$ from this FC layer is then passed through a sigmoid layer and normalized by,
\begin{equation}
p_{i}^{I}=\frac{1}{1 + \exp{\left(-f_{i}^{I}\right)}}
\end{equation}
where $I$ is a CXR image and $p_{i}^{I}$ represents the probability score of $I$ belonging to the $i^{th}$ class, where $i \in \{1, 2,\ldots, n\}$. $n$ represents the number of pathologies presented in each dataset.
\subsubsection{\textbf{Loss functions}}
The pathological labels of each CXR are expressed as an $n$-dimensional label vector, $L=[l_1, \ldots, l_i, \ldots, l_n]$, where $l_i \in \{0, 1\}$. $l_i$ denotes whether there is any pathology, i.e., 1 for presence and 0 for absence. We employ binary cross-entropy loss for optimization, defined by:
\begin{equation}
\mathcal{L}_{cls}= -\frac{1}{n}\sum_{i=1}^{n}\bigg[l_i\log{\left(p_{i}^{I}\right)}+(1-l_i)\log{\left(1-p_{i}^{I}\right)}\bigg] 
\end{equation}

\begin{table*}[!t]
    \centering
    \caption{\textsc{Pathology-wise Performance Comparison of the Proposed Method with State-of-the-art Systems on the NIH Dataset $^a$. The Two Best Results are Shown in \textcolor{red}{Red} and \textcolor{blue}{Blue}.}}
    \label{nih_comparison}
    \begin{adjustbox}{width=\textwidth}
    \begin{threeparttable}[b]
    \begin{tabular}{c|cccccccccccccc|c} 
    \hline
    \toprule 
    \bf Method & \bf Emph & \bf Fibr & \bf Hern & \bf Infi & \bf PT & \bf Mass & \bf Nodu & \bf Atel & \bf Card & \bf Cons & \bf Edem & \bf Effu & \bf Pne1 & \bf Pne2 & \bf Average \\
    \hline\hline
    \multicolumn{16}{c}{Methods without Utilizing Segmentation Masks} \\
    \hline
    Ho \etal \cite{ho2019multiple} & 87.50 & 75.60 & 83.60 & 70.30 & 77.40 & 83.50 & 71.60 & 79.50 & 88.70 & 78.60 & 89.20 & 87.50 & 74.20 & 86.30 & 80.97 \\
    CRAL \cite{GUAN2020259} & 90.80 & 83.00 & 91.70 & 70.20 & 77.80 & 83.40 & 77.30 & 78.10 & 88.00 & 75.40 & 85.00 & 82.90 & 72.90 & 85.70 & 81.59 \\
    CheXNet \cite{Rajpurkar2017CheXNetRP} & 92.49 & 82.19 & 93.23 & 68.94 & 79.25 & 83.07 & 78.14 & 77.95 & 88.16 & 75.42 & 84.96 & 82.68 & 73.54 & 85.13 & 81.80 \\
    DualCheXNet \cite{chen2019dualchexnet} & 94.20 & 83.70 & 91.20 & 70.50 & 79.60 & 83.80 & 79.60 & 78.40 & 88.80 & 74.60 & 85.20 & 83.10 & 72.70 & 87.60 & 82.30 \\
    LLAGNet \cite{Chen2020LesionLA} & 93.90 & 83.20 & 91.60 & 70.30 & 79.80 & 84.10 & 79.00 & 78.30 & 88.50 & 75.40 & 85.10 & 83.40 & 72.90 & 87.70 & 82.37 \\
    Wang \etal \cite{wang2020triple} & 93.30 & 83.80 & 93.80 & 71.00 & 79.10 & 83.40 & 77.70 & 77.90 & 89.50 & 75.90 & 85.50 & 83.60 & 73.70 & 87.80 & 82.60 \\
    Yan \etal \cite{Yan2018WeaklySD} & \textcolor{blue}{\bf 94.22} & 83.26 & 93.41 & 70.95 & \textcolor{red}{\bf 80.83} & 84.70 & \textcolor{blue}{\bf 81.05} & 79.24 & 88.14 & 75.98 & 84.70 & 84.15 & 73.97 & 87.59 & 83.02 \\
    Luo \etal \cite{Luo2020DeepME} & 93.96 & 83.81 & 93.71 & \textcolor{blue}{\bf 71.84} & \textcolor{blue}{\bf 80.36} & 83.76 & 79.85 & 78.91 & 90.69 & 76.81 & 86.10 & 84.18 & 74.19 & \textcolor{red}{\bf 90.63} & 83.49 \\
    \hline
    \multicolumn{16}{c}{Methods Utilizing Segmentation Masks} \\
    \hline
    Arias-Garzón \etal \cite{ARIASGARZON2021100138} & 85.72 & 81.68 & 82.48 & 70.10 & 77.67 & 83.63 & 78.92 & 80.43 & 88.93 & 80.17 & 87.71 & 86.89 & 75.07 & 85.59 & 81.79 \\
    MANet \cite{XU202196} & 85.23 & 82.82 & 92.10 & 70.04 & 76.82 & 83.36 & 77.76 & 81.43 & 89.35 & 80.23 & 88.56 & 86.30 & 75.29 & 85.46 & 82.48 \\
    Keidar \etal \cite{Keidar2021} & 90.87 & 81.47 & 91.80 & 70.60 & 78.02 & 83.93 & 77.07 & 80.64 & 90.88 & 80.43 & 89.20 & 86.94 & 76.53 & 85.54 & 83.14 \\
    Anatomy-XNet (224) & 92.85 & \textcolor{blue}{\bf 84.42} & \textcolor{red}{\bf 96.36} & 71.71
    & 79.79 & \textcolor{blue}{\bf 86.04} & 80.37 & \textcolor{blue}{\bf 83.06} & \textcolor{blue}{\bf 91.37} & \textcolor{blue}{\bf 80.91} & \textcolor{blue}{\bf 89.90} & \textcolor{blue}{\bf 88.58} & \textcolor{blue}{\bf 77.09} & 88.21 & \textcolor{blue}{\bf 85.05} \\
    Anatomy-XNet (512) & \textcolor{red}{\bf 94.33} & \textcolor{red}{\bf 85.91} & \textcolor{blue}{\bf 94.57} & \textcolor{red}{\bf 72.07}
    & 79.90 & \textcolor{red}{\bf 86.80} & \textcolor{red}{\bf 83.78} & \textcolor{red}{\bf 83.69} & \textcolor{red}{\bf 91.38} & \textcolor{red}{\bf 81.54} & \textcolor{red}{\bf 90.25} & \textcolor{red}{\bf 89.12} & \textcolor{red}{\bf 77.48} & \textcolor{blue}{\bf 90.09} & \textcolor{red}{\bf 85.78} \\ 
    \bottomrule 
    \end{tabular}
    \begin{tablenotes}
    \item[a] The 14 findings for NIH datasets are Emphysema (Emph), Fibrosis (Fibr), Hernia (Hern), Infiltration (Infi), Pleural Thickening (PT), Mass, Nodule (Nodu), Atelectasis (Atel), Cardiomegaly (Card), Consolidation (Cons), Edema (Edem), Effusion (Effu), Pneumonia (Pne1), and Pneumothorax (Pne2).
    \end{tablenotes}
    \end{threeparttable}
    \end{adjustbox}
\end{table*}

\begin{table*}[!t]
    \centering
    \caption{\textsc{Pathology-wise Performance Comparison of the Proposed Method with State-of-the-art Systems on the CheXpert Dataset. 
    The Two Best Results are Shown in \textcolor{red}{Red} and \textcolor{blue}{Blue}.}}
    \label{chexpert_comparison}
    \begin{tabular}{c|ccccc|c} 
    \toprule
    \bf Method & \bf Atelectasis & \bf Cardiomegaly & \bf Edema & \bf Consolidation & \bf Pleural Effusion & \bf Average \\
    \hline\hline
    \multicolumn{7}{c}{Methods without Utilizing Segmentation Masks} \\
    \hline
    Allaouzi \etal \cite{8719904} BR & 72.00 & \textcolor{blue}{\bf 88.00} & 87.00 & 77.00 & 90.00 & 82.80 \\
    Irvin \etal \cite{irvin2019chexpert} U-Ones & 85.80 & 83.20 & 94.10 & 89.90 & 93.40 & 89.30 \\
    Pham \etal \cite{PHAM2021186} U-Ones+CT+LSR & 82.50 & 85.50 & 93.00 & \textcolor{red}{\bf 93.70} & 92.30 & 89.40 \\
    \hline
    \multicolumn{7}{c}{Methods Utilizing Segmentation Masks} \\
    \hline
    MANet \cite{XU202196} & 81.35 & 86.61 &	92.22 &	91.59 & 89.86	& 88.33 \\
    Arias-Garzón \etal \cite{ARIASGARZON2021100138} & 81.74	& 84.24 & 94.06 & 90.74 & 94.31 & 89.02 \\
    Keidar \etal \cite{Keidar2021} & 86.42 & 87.39 & 91.97 & 88.23 & 91.73 & 89.15 \\
    Anatomy-XNet (224) & \textcolor{blue}{\bf 86.55} & 87.86 & \textcolor{blue}{\bf 95.28} & 93.13 & \textcolor{blue}{\bf 94.66} & \textcolor{blue}{\bf 91.50} \\
    Anatomy-XNet (512) & \textcolor{red}{\bf 86.72} & \textcolor{red}{\bf 89.54} & \textcolor{red}{\bf 95.73} & \textcolor{blue}{\bf 93.31} & \textcolor{red}{\bf 95.04} & \textcolor{red}{\bf 92.07} \\
    \bottomrule 
    \end{tabular}
\end{table*}

\begin{table*}[!t]
    \centering
    \caption{\textsc{Pathology-wise Performance Comparison of the Proposed Method with State-of-the-art Systems on the MIMIC-CXR Dataset $^b$. The Two Best Results are Shown in \textcolor{red}{Red} and \textcolor{blue}{Blue}.}}
    \label{mimic_comparison}
    \begin{adjustbox}{width=\textwidth}
    \begin{threeparttable}[b]
    \begin{tabular}{c|cccccccccccccc|c} 
    \hline
    \toprule 
    \bf Method & \bf Atel & \bf Card & \bf Cons & \bf Edem & \bf E.C. & \bf Frac & \bf L.L. & \bf L.O. & \bf N.F. & \bf Effu & \bf P.O. & \bf Pne1 & \bf Pne2 & \bf S.D. & \bf Average \\
    \hline\hline
    \multicolumn{16}{c}{Methods without Utilizing Segmentation Masks} \\
    \hline
    Densenet-KG \cite{Zhang_Wang_Xu_Yu_Yuille_Xu_2020} & 69.40 & 74.60 & 64.00 & 79.00 & 65.10 & 60.50 & 57.40 & 60.90 & 77.80 & 80.90 & 65.00 & 57.20 & 68.90 & 78.10 & 68.50 \\
    VSE-GCN \cite{Hou2021MultiLabelLW} & 72.20 & 73.00 & 72.80 & 79.90 & 76.70 & 56.00 & 62.30 & 65.40 & 81.70 & 86.30 & 65.30 & 58.80 & 79.70 & 78.90 & 72.10 \\ 
    Chexclusion \cite{SeyyedKalantari2021CheXclusionFG} $^c$ & 83.70 & \textcolor{red}{\bf 82.80} & 84.40 & 90.40 & \textcolor{red}{\bf 75.70} & 71.80 & \textcolor{blue}{\bf 77.20} & 78.20 & 86.80 & 93.30 & 84.80 & 74.80 & 90.30 & 92.70 & 83.40 \\
    \hline
    \multicolumn{16}{c}{Methods Utilizing Segmentation Masks} \\
    \hline
    Arias-Garzón \etal \cite{ARIASGARZON2021100138} & 82.61 & 81.57 & 83.16 & 90.01 & 73.71 & 65.36 & 74.57  & 77.40 & 85.83 & 91.50 & 81.96 & 72.79 & 87.47  & 90.64 & 81.33 \\
    MANet \cite{XU202196} & 82.77 & 81.86 & 83.66 & 90.03 & 74.52 & 69.56 & 75.43 & 77.24 & 85.90 & 91.53 & 83.05 & 73.01 & 88.02 & 90.24 & 81.92 \\
    Keidar \etal \cite{Keidar2021} & 83.24 & 82.59 & 84.19 & 90.40 & 74.71 & 71.33 & 76.66 & 77.67 & 86.39 & 92.93 & 84.18 & 74.51 & 89.70 & 92.05 & 82.90 \\
    Anatomy-XNet (224) & \textcolor{blue}{\bf 83.79} & \textcolor{blue}{\bf 82.67} & \textcolor{red}{\bf 85.25} & \textcolor{red}{\bf 90.83} & \textcolor{blue}{\bf 75.45} & \textcolor{blue}{\bf 74.30} & 77.08 & \textcolor{blue}{\bf 78.79} & \textcolor{blue}{\bf 86.90} & \textcolor{blue}{\bf 93.37} & \textcolor{red}{\bf 86.55} & \textcolor{red}{\bf 75.98} & \textcolor{blue}{\bf 90.87} & \textcolor{blue}{\bf 92.75} & \textcolor{blue}{\bf 83.90} \\ 
    Anatomy-XNet (512) & \textcolor{red}{\bf 83.93} & 82.59 & \textcolor{blue}{\bf 84.84} & \textcolor{blue}{\bf 90.76} & 75.12 & \textcolor{red}{\bf 74.95} & \textcolor{red}{\bf 78.78} & \textcolor{red}{\bf 78.90} & \textcolor{red}{\bf 86.97} & \textcolor{red}{\bf 93.43} & \textcolor{blue}{\bf 86.21} & \textcolor{blue}{\bf 75.81} & \textcolor{red}{\bf 91.20} & \textcolor{red}{\bf 93.12} & \textcolor{red}{\bf 84.04} \\
    
    \bottomrule 
    \end{tabular}
    \begin{tablenotes}
    \item[b] The 14 pathologies for the MIMIC-CXR datasets are Atelectasis (Atel), Cardiomegaly (Card), Consolidation (Cons), Edema (Edem), Enlarged Cardiomediastinum (E.C.), Fracture (Frac), Lung Lesion (L.L.), Lung Opacity (L.O.), No Finding (N.F.), Pleural Effusion (Effu), Pleural Other (P.O.), Pneumonia (Pne1), Pneumothorax (Pne2), Support Devices (S.D.).
    \item[c] Indicates that the result is obtained by the ensemble of 5 checkpoints.
    \end{tablenotes}
    \end{threeparttable}
    \end{adjustbox}
\end{table*}

\section{Training}
\subsection{Datasets}
\textbf{NIH}: The NIH chest X-ray dataset \cite{8099852} consists of 112,120 X-rays from 30,805 unique patients with 14 diseases. We strictly follow the official split of NIH, 70\% for training, 10\% for validation, and 20\% for testing, for conducting experiments and fair comparison with previous works. \par
\textbf{CheXpert}: The CheXpert dataset \cite{irvin2019chexpert} consists of 224,316 X-rays of 65,240 patients. The official specific validation and test datasets consist of 200, and 500 studies respectively. \par 
\textbf{MIMIC-CXR}: The MIMIC-CXR dataset \cite{johnson2019mimiccxrjpg} contains 377,111 X-rays with 14 diseases. We combine all non-positive labels (negative, not mentioned, and uncertain) into an aggregate negative label \cite{SeyyedKalantari2021CheXclusionFG} for experimenting on this dataset. \par
\textbf{JSRT}: We use the JSRT \cite{Shiraishi2000DevelopmentOA} as annotated dataset to train the segmentation model. The segmentation annotations for JSRT, including heart and lung, are obtained from \cite{VANGINNEKEN200619}. 

\subsection{Implementation Details}
\subsubsection{Training Scheme for Segmentation}
\label{training_scheme: segmentation}
We follow the procedure outlined in \cite{Mondal2019RevisitingCF} for preprocessing, output binarization, and hyper-parameter settings to utilize the semi-supervised training pipeline. We utilize the large-scale datasets as the unannotated subset, i.e., for training the model on the NIH dataset, the NIH dataset is used as an unannotated subset. Thus, we get three separate segmentation models, where NIH, CheXpert, and MIMIC-CXR datasets are used as an unannotated subset, respectively. We use the JSRT dataset as the annotated subset in all three cases. Calculating accuracy in external datasets such as NIH, CheXpert, or MIMIC-CXR is impossible due to the unavailability of the ground truths for them. For this reason, the validation dataset for the semi-supervised setting in all three cases is comprised of CXR images from the JSRT dataset. We choose the checkpoint with the highest dice score on this validation dataset as the final model. 
\begin{figure*}[!t]
    \centering
	\includegraphics[width=0.9\linewidth]{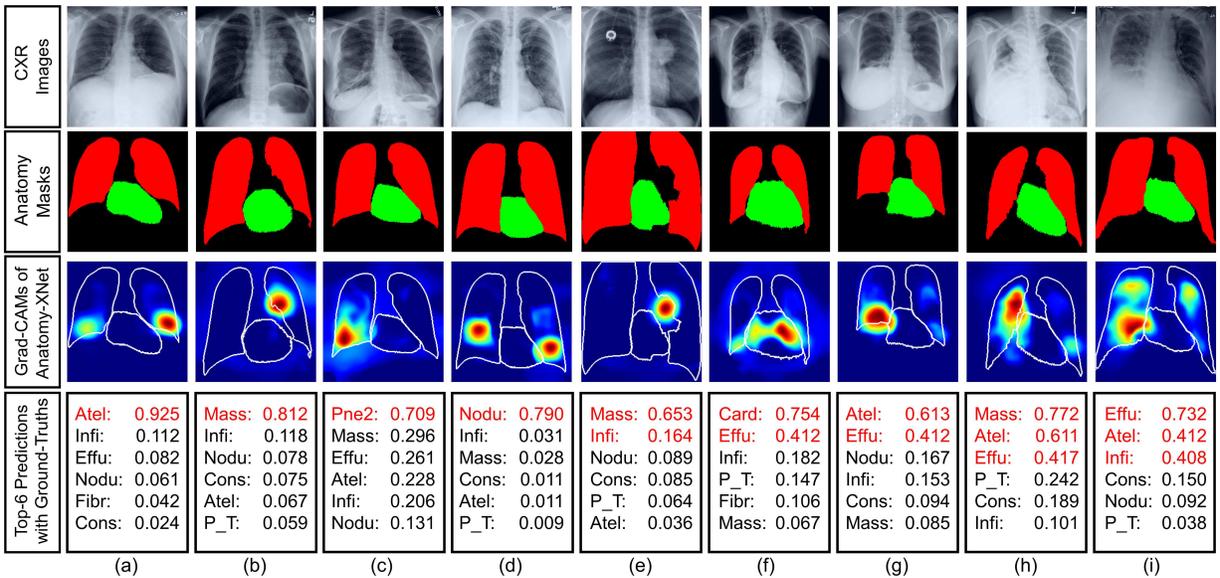}
    \caption{Qualitative visualization of the Anatomy-XNet's output on the NIH test dataset. The first and second rows depict the CXR images and their corresponding anatomy masks predicted from the semi-supervised segmentation network, respectively. The color red in the segmentation masks indicates the lung regions, while the color green indicates the heart regions. The third row shows the Grad-CAMs of the Anatomy-XNet for the target classes. The color red in the Grad-CAMs means the most indicative regions with abnormalities, while the color blue indicates regions without abnormalities. The contours of the anatomy segmentation masks are marked with white color on top of the heatmaps. The final row shows top-6 predicted findings and their corresponding prediction scores. The ground truth labels are highlighted in \textcolor{red}{red} color.}
    \label{attention_figure}
\end{figure*}
\copyrightnotice
\subsubsection{Training Scheme for Classification}
In terms of image size, for a fair comparison with others, we follow \cite{wang2020triple, Rajpurkar2017CheXNetRP, XU202196, PHAM2021186} and resize the CXR images to 256\texttimes256, and then randomly crop 224\texttimes224 patches as inputs \cite{Yan2018WeaklySD}. However, 224\texttimes224 is too small to predict small and subtle diseases like nodules, pneumonia in CXR. For that reason, we train our model on a larger image size as well. To utilize the larger input image dimension, we resize the CXR images to 586\texttimes586, and then randomly crop 512\texttimes512 patches as inputs. We normalize the input images with the mean and standard deviation of the ImageNet training set. We follow \cite{Yan2018WeaklySD} and take advantage of flipping to increase the variation and the diversity of training samples. For validation and inference, we use a centrally cropped sub-image of 512\texttimes512 for 586\texttimes586 and 224\texttimes224 for 256\texttimes256 dimensions as input. The anatomy masks from the segmentation network are resized to 56\texttimes56 before passing to the A\textsuperscript{3} modules. We use Adam optimizer with an initial learning rate of 0.0001 and set the batch size to 120. Following previous studies, \cite{8099852, Luo2020DeepME}, we employ the percentage area under the receiver operating characteristic curve (AUC) for performance evaluation.

\section{Experimental Results And Analysis}
\subsection{Comparison With State-of-the-Arts}
\subsubsection{Performance on NIH-dataset}
We compare our proposed Anatomy-XNet with previously published state-of-the-art methods including: Category-wise Residual Attention Learning (CRAL) \cite{GUAN2020259}, CheXNet \cite{Rajpurkar2017CheXNetRP}, DualCheXNet \cite{chen2019dualchexnet}, Lesion Location Attention Guided Network (LLAGNet) \cite{Chen2020LesionLA}, the methods of Ho \etal \cite{ho2019multiple}, Wan \etal \cite{wang2020triple}, Yan \etal \cite{Yan2018WeaklySD}, Luo \etal \cite{Luo2020DeepME}, Arias-Garzón \etal \cite{ARIASGARZON2021100138}, Keidar \etal \cite{Keidar2021}, and MANet \cite{XU202196}. We have implemented the methods of Arias-Garzón \etal \cite{ARIASGARZON2021100138}, Keidar \etal \cite{Keidar2021}, and MANet \cite{XU202196}. For the methods of Ho \etal \cite{ho2019multiple} and DualCheXNet \cite{chen2019dualchexnet}, results have been reported from their implementations. The results for the rest of the methods are quoted from \cite{Luo2020DeepME}. As shown in Table~\ref{nih_comparison}, the method proposed by Luo \etal \cite{Luo2020DeepME} is the previous state-of-the-art yielding an AUC of 83.49\%, while our proposed Anatomy-XNet exceeds all the compared models and achieves a new state-of-the-art performance of 85.05\% AUC. With a higher input image dimension of 512\texttimes512, our proposed framework boosts performance to an AUC score of 85.78\%. Specifically, our classification results outperform others in 12 out of 14 categories.

\subsubsection{Performance on CheXpert-dataset}
The results on CheXpert are compared in Table~\ref{chexpert_comparison}. In this paper, we focus on comparing the results achieved by a single model architecture. We quote the single model performance for Pham \etal \cite{PHAM2021186}, and Allaouzi \etal \cite{8719904} from their implementations. We report the ensemble result of Irvin \etal \cite{irvin2019chexpert} as single checkpoint performance is not given in their paper. To compare with approaches that have utilized segmentation masks, we have implemented the methods of Arias-Garzón \etal \cite{ARIASGARZON2021100138}, Keidar \etal \cite{Keidar2021}, and MANet \cite{XU202196}. The results from Table~\ref{chexpert_comparison} show that our model achieves an AUC of 91.50\% and 92.07\% with input image dimensions of 224\texttimes224 and 512\texttimes512, respectively, surpassing the previous state-of-the-art results.
\copyrightnotice
\subsubsection{Performance on MIMIC-CXR-dataset}
We compare our proposed Anatomy-XNet with previously published state-of-the-art methods including: Densenet-KG \cite{Zhang_Wang_Xu_Yu_Yuille_Xu_2020}, VSE-GCN \cite{Hou2021MultiLabelLW}, CheXclusion \cite{SeyyedKalantari2021CheXclusionFG}, the methods of Keidar \etal \cite{Keidar2021}, MANet \cite{XU202196}, and Arias-Garzón \etal \cite{ARIASGARZON2021100138}. We adopt the same data split procedure outlined in \cite{SeyyedKalantari2021CheXclusionFG}. The results of Densenet-KG \cite{Zhang_Wang_Xu_Yu_Yuille_Xu_2020}, VSE-GCN \cite{Hou2021MultiLabelLW} have been quoted from the implementation of VSE-GCN \cite{Hou2021MultiLabelLW}. We report the result of  CheXclusion \cite{SeyyedKalantari2021CheXclusionFG} from their implementation. We have implemented the methods of Arias-Garzón \etal \cite{ARIASGARZON2021100138}, Keidar \etal \cite{Keidar2021}, and MANet \cite{XU202196}. The results are shown in Table~\ref{mimic_comparison}. Our proposed model with both input image dimensions of 224\texttimes224 and 512\texttimes512 has achieved higher performance than the compared models.

\subsection{Impact of Semi-supervised Segmentation}
\begin{figure}[t]
    \centering
	\includegraphics[width=0.7\linewidth]{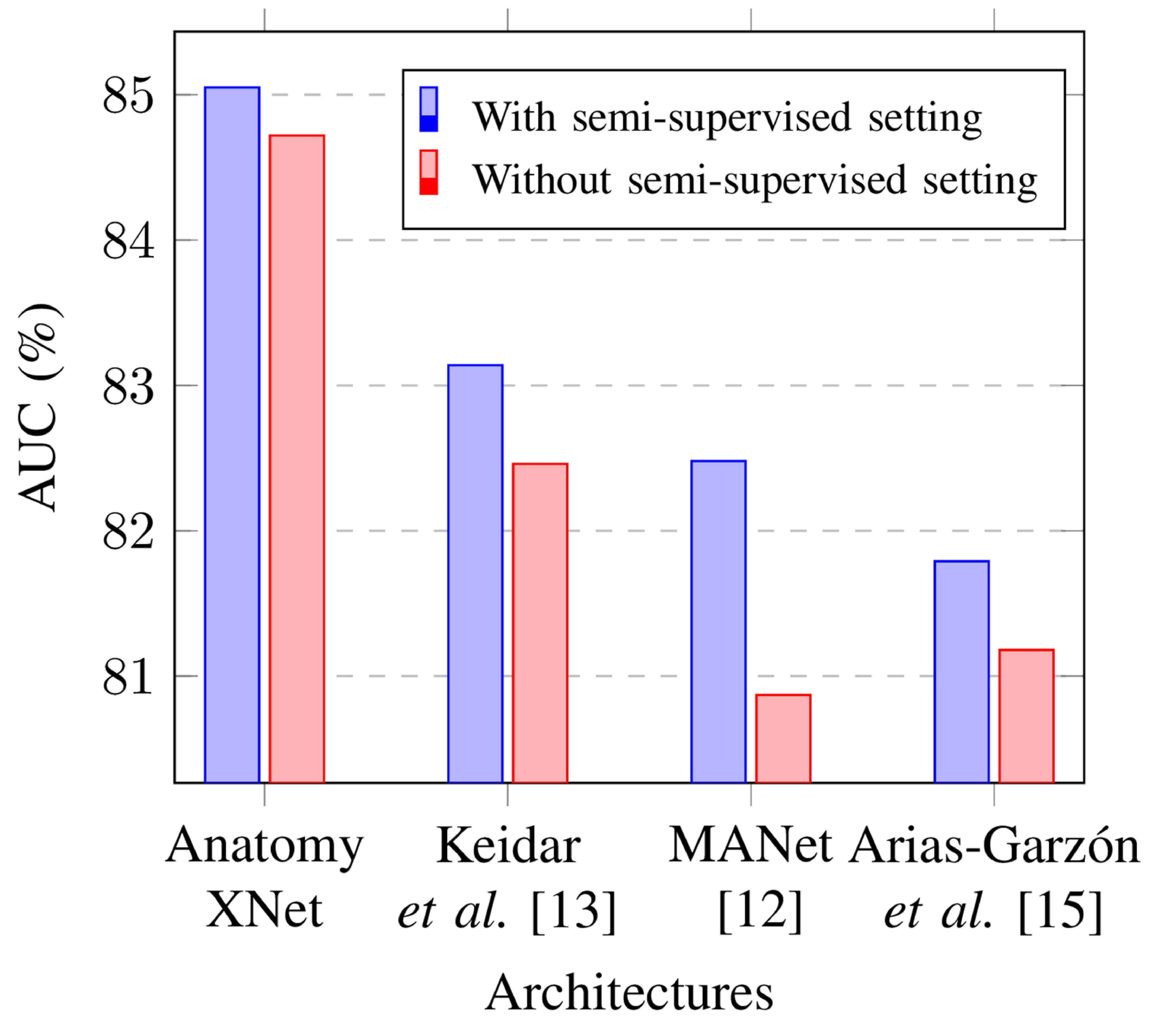}
    \caption{Investigation of the performance of the segmentation-mask-based classification methods, with masks generated with the semi-supervised segmentation setting compared to masks generated without the semi-supervised segmentation setting.}
    \label{fig: segmentation mask impact comparison bar plot}
\end{figure}

The selected models for NIH, CheXpert, and MIMIC-CXR datasets achieve validation dice scores of 0.7437, 0.7395, and 0.7417, respectively. These models are used to generate anatomy masks for their corresponding datasets. The visualizations of predicted segmentation results on the NIH test dataset are given in the second row of Fig.~\ref{attention_figure}. To verify the impact of the quality of the semi-supervised segmentation masks on the classification performance, we train the segmentation network, only on the labeled dataset (JSRT), without the semi-supervised setting. Next, we train the segmentation-based methods \cite{Keidar2021, XU202196, ARIASGARZON2021100138}, including Anatomy-XNet, on the NIH dataset with masks generated from this segmentation network (without the semi-supervised setting), and measure their performance on the NIH test dataset. The results are reported in Fig.~\ref{fig: segmentation mask impact comparison bar plot}, where we observe that the classification performance of all the methods improves by utilizing masks generated from the semi-supervised setting. In addition, we also observe that the drop in performances of other methods \cite{Keidar2021, XU202196, ARIASGARZON2021100138} is larger compared to Anatomy-XNet due to their lack of robustness to counter imperfection in segmentation masks.

\subsection{Qualitative Visualization and Analysis}
\label{attention_region_attention_vector_analysis}
We generate attention heatmaps using Gradient-weighted Class Activation Mappings (Grad-CAMs) \cite{8237336} to visualize the most indicative pathology areas on CXRs from the NIH test dataset to interpret the representational power of Anatomy-XNet. These attention heatmaps, along with the CXRs, anatomy masks predicted from the semi-supervised segmentation network, and classification results, are shown in Fig.~\ref{attention_figure}. A visual evaluation of the Grad-CAMs confirms the module's anatomy awareness. Thus, similar to the process followed by a radiologist, the A\textsuperscript{3} module integrates the anatomy information responsible for a particular pathology within the model. In cases of imperfect mask segmentation (due to semi-supervised training setting), our proposed method still manages to capture the pathology relevant areas and give attention to them. Column (e) of Fig.~\ref{attention_figure} demonstrates an example where the lung mask fails to contain the mass area. Nevertheless, our model localizes its attention in that area, demonstrating the efficacy of the proposed architecture's resilience towards imperfect segmentation.

\begin{table}[!t]
    \centering
    \caption{\textsc{Investigation of the Classification Performance in the NIH, MIMIC-CXR, and CheXpert Datasets with Different Settings. The Best Result is Shown in \textcolor{red}{Red}.}}
    \label{all_ablation}
    \begin{tabular}{ccccc}
    \toprule
    \multicolumn{5}{c}{\textbf{Part 1: Investigation of the effectiveness of A\textsuperscript{3} modules.}} \\
    \hline
    Dataset & Baseline & A\textsuperscript{3}-L1 & A\textsuperscript{3}-L2 & A\textsuperscript{3}-L3 \\
    \hline
    NIH & 82.44 & 84.67 & \textcolor{red}{\bf 85.05} & 84.72 \\
    MIMIC-CXR & 82.76 & 83.72 & \textcolor{red}{\bf 83.90} & 83.67 \\
    CheXpert & 89.22 & 90.93 & \textcolor{red}{\bf 91.50} & 91.07 \\
    \hline
    \multicolumn{5}{c}{\textbf{Part 2: Investigation of the effectiveness of PWAP modules.}}\\
    \hline
    Dataset & PWAP & Gem & Average & Max \\
    \hline
    NIH & \textcolor{red}{\bf 85.05} & 84.85 &  84.45 & 84.30 \\
    MIMIC-CXR & \textcolor{red}{\bf 83.90} & 83.68 & 83.46 & 83.32 \\
    CheXpert & \textcolor{red}{\bf 91.50} & 91.18 & 91.03 & 90.42 \\
    \hline
    \multicolumn{5}{c}{\textbf{Part 3: Investigation of different anatomy mask sizes.}} \\
    \hline
    Dataset & 28\texttimes28 & 42\texttimes42 & 56\texttimes56 & - \\
    \hline
    NIH & 84.86 & 84.90 & \textcolor{red}{\bf 85.05} & - \\
    MIMIC-CXR & 83.38 & 83.71 & \textcolor{red}{\bf 83.90} & - \\
    CheXpert & 91.21 & 91.21 & \textcolor{red}{\bf 91.50} & - \\
    \hline 
    \multicolumn{5}{c}{\textbf{Part 4: Investigation of different input image sizes.}} \\
    \hline
    Dataset & 224\texttimes224 & 384\texttimes384 & 512\texttimes512 & - \\
    \hline
    NIH & 85.05 & 85.44 & \textcolor{red}{\bf 85.78} & - \\  
    MIMIC-CXR & 83.90 & 83.97 & \textcolor{red}{\bf 84.04} & - \\ 
    CheXpert & 91.50 & 91.70 & \textcolor{red}{\bf 92.07} & - \\ 
    \bottomrule
    \end{tabular}
\end{table}
\copyrightnotice
\subsection{Effectiveness of \texorpdfstring{A\textsuperscript{3}}{A3} Modules}
For evaluating the impact of A\textsuperscript{3} modules on classification performance, we cascade multiple A\textsuperscript{3} modules with different dense blocks (DB). First, we use an A\textsuperscript{3} module with DB-4. We denote this experiment by anatomy aware attention level-1 (A\textsuperscript{3}-L1). Afterward, we use A\textsuperscript{3} modules with DB-3,4 and indicate this by anatomy aware attention level-2 (A\textsuperscript{3}-L2). Finally, we apply the A\textsuperscript{3} modules with DB-2,3,4 and refer to it as anatomy aware attention level-3 (A\textsuperscript{3}-L3). The experimental results are provided in part-1 of Table~\ref{all_ablation}. Our experiments find that classification performance improves from the baseline when we cascade a A\textsuperscript{3} module with a DB. The baseline denotes the backbone model, DenseNet-121, without any integrated A\textsuperscript{3} modules. The results show that performance improves when going from A\textsuperscript{3}-L1 to A\textsuperscript{3}-L2 but decreases if A\textsuperscript{3}-L3 is used. Because low-level spatial features from DB-2 might have outlier information which deteriorates the performance by causing the model to give attention to noisy information. Again, applying A\textsuperscript{3} only on the highest level of features, in our case DB-4, does not guarantee the best performance. Because due to subsequent pooling in these DBs, some salient information presented in the previous DBs, may be lost in the later stages. 

\subsection{Effectiveness of PWAP Modules}
\begin{figure}[t]
    \centering
	\includegraphics[width=0.85\linewidth]{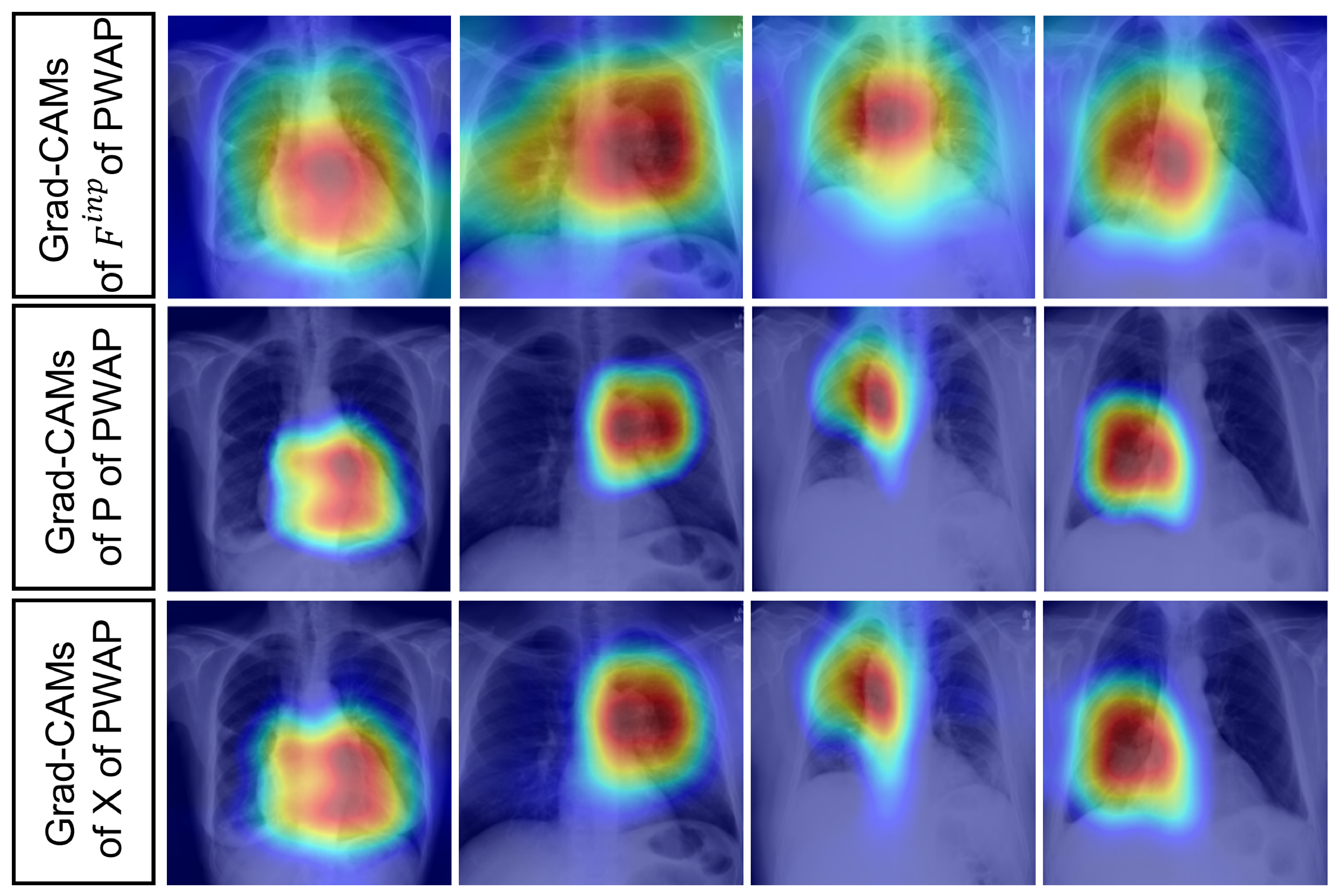}
    \caption{Visualization of the impact of the PWAP module. The first, second, and third row depict the Grad-CAMs of the input feature spaces ($\mathbf{F}^{inp}$), probability attention maps ($\mathbf{P}$), and recalibrated feature spaces ($\mathbf{X}$) of the PWAP module, which is inside the A\textsuperscript{3} module connected with the fourth dense block.}
\label{pwap_output}
\end{figure}

To demonstrate the effectiveness of PWAP modules, we replace all the PWAP layers in Anatomy-XNet with average pooling, max pooling, and generalized mean pooling \cite{Berman2019MultiGrainAU} layers, respectively, and run the experiment while keeping all the other hyperparameters the same. The results across all the datasets, shown in part-2 of Table~\ref{all_ablation}, depict the effectiveness of the proposed module. In Fig.~\ref{pwap_output}, the Grad-CAMs of the input feature spaces ($\mathbf{F}^{inp}$), probability attention maps ($\mathbf{P}$), and recalibrated feature spaces ($\mathbf{X}$) of the PWAP module, that is inside the A\textsuperscript{3} module connected with the fourth dense block, are shown. The heatmaps are resized to the dimensions of the CXR images and overlaid on the CXR images. A visual examination of the probability attention maps and the recalibrated feature space shows that the PWAP module modulates the feature space to focus more prominently on the lesion areas by removing unwanted attention.

\subsection{Effect of Different Anatomy Mask Dimensions}
To demonstrate the effect of the dimension of anatomy masks, we vary the intermediate dimensions of the anatomy masks, chosen from the set \{28\texttimes28, 42\texttimes42, 56\texttimes56\}, and evaluate the performances of Anatomy-XNet. Part-3 of Table~\ref{all_ablation} presents performance numbers across all datasets. Here, we observe that the performance of Anatomy-XNet improves as we increase the dimension of the anatomy masks. 

\subsection{Effect of Different Input Image Sizes}
We perform experiments to investigate the effect of varying input image dimensions on classification performance. We resize the CXR images into three different sizes: 256\texttimes256, 438\texttimes438, and 586\texttimes586 and crop patches of 224\texttimes224 for 256\texttimes256, 384\texttimes384 for 438\texttimes438, and 512\texttimes512 for 586\texttimes586 to use as input images. The classification performances on all three datasets for different input sizes are given in the part-4 of Table~\ref{all_ablation}. The classification results show that enlarging the input image size increases the average AUC.
\copyrightnotice
\subsection{Investigation of the Impact of Imperfect Segmentation}
\begin{figure}[t]
    \centering
	\includegraphics[width=0.9\linewidth]{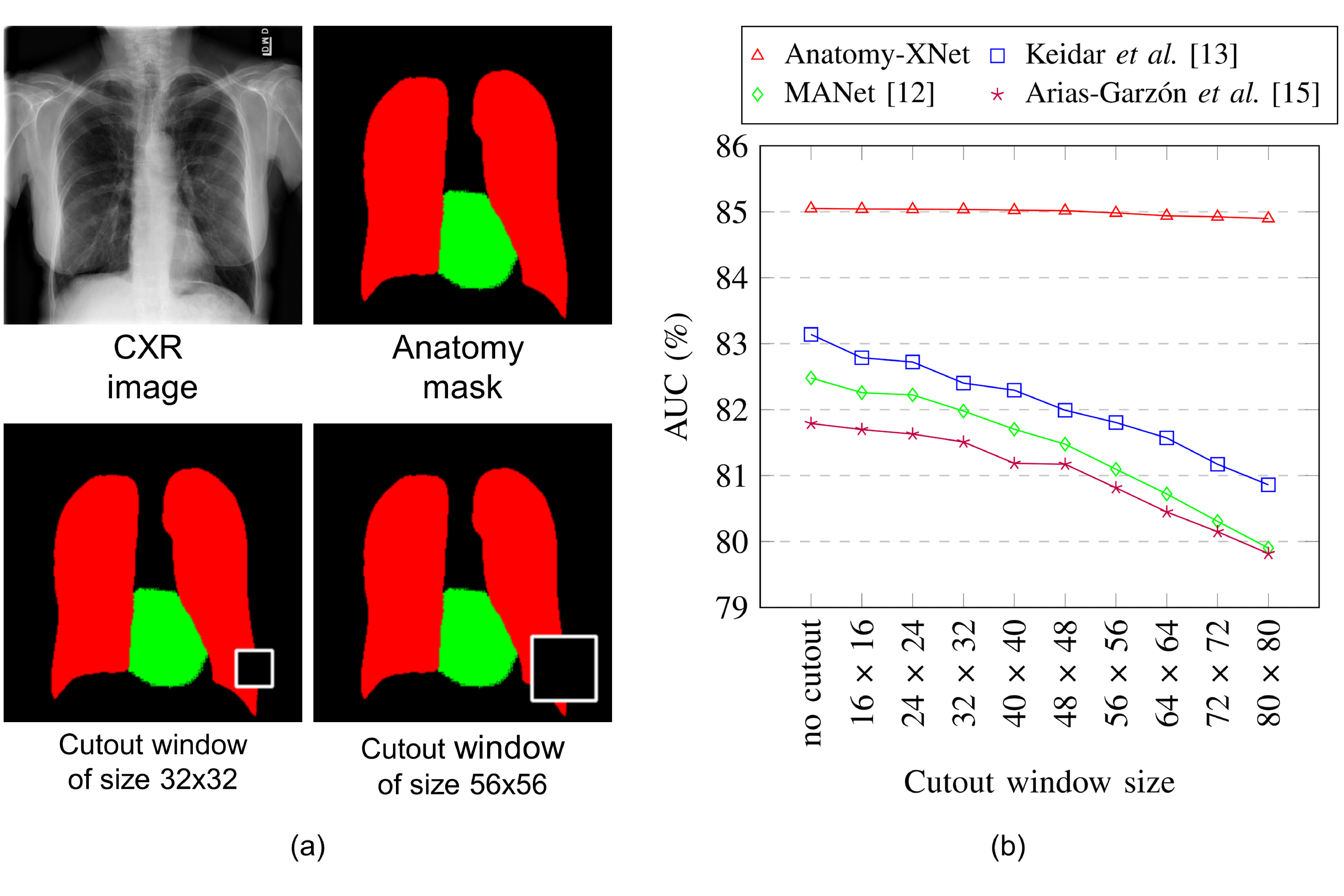}
    \caption{(a) Visualization of cutout window of two different sizes. (b) Simulation of classification performance drop against increasing imperfection in segmentation masks.}
    \label{performance_drop_simulation_nih}
\end{figure}

To simulate the resilience of the proposed Anatomy-XNet towards imperfect segmentation masks, we randomly apply cutout operations\cite{Devries2017ImprovedRO} on the predicted anatomy segmentation regions with different window sizes and measure the AUC score. The NIH, CheXpert, and MIMIC-CXR datasets do not contain pixel-level ground truth annotations. Due to the lack of pixel-level ground truth annotations and the sheer size of the datasets, it is very challenging to ensure that the cutout window will always be on the lesion area. Instead, we make sure that the regions on which the cutout windows are applied always overlap with the predicted anatomy masks. We perform the cutout operation three times for each window size, measure AUC each time, and take the average as the final AUC score for that particular cutout window. Next, we apply the exact same cutout operations at exactly the same locations of the anatomy masks and use them to evaluate the AUC of segmentation mask-based approaches \cite{XU202196, Keidar2021, ARIASGARZON2021100138} and compare their drop in classification performance with our method. The AUC scores against different cutout window sizes are shown in Fig.~\ref{performance_drop_simulation_nih}. The proposed Anatomy-XNet shows only around 0.2\% performance degradation against cutout operations and maintains stable performance against increasing window sizes. On the other hand, the methods of \cite{Keidar2021, ARIASGARZON2021100138, XU202196} show a larger degradation in classification performance, around 2.41-3.13\%, against increasing cutout window size.
\section{Conclusion}
In this paper, we propose Anatomy-XNet, an anatomy-aware convolutional neural network for thoracic disease classification. Departing from the previous works that rely on the chest X-ray image only or attention mechanisms guided by the model prediction, the proposed network is guided by prior anatomy segmentation information to act similar to a radiologist by focusing on relevant anatomical regions associated with the thoracic disease. Extensive experiments demonstrate that combining our novel A\textsuperscript{3} and PWAP modules within a backbone Densenet-121 model in a unified framework yields state-of-the-art performance on the NIH chest X-ray, Stanford CheXpert, and  MIMIC-CXR datasets. The Anatomy-XNet achieves an average AUC score of 85.78\% on the official NIH test set, 92.07\% on the official validation split of the Stanford CheXpert dataset, and 84.04\% on the MIMIC-CXR dataset, surpassing the former best-performing methods published on these datasets.
\copyrightnotice
\balance
\bibliographystyle{IEEEtran}
\bibliography{IEEEabrv,main.bib}

\end{document}